\documentclass[twocolumn,A4]{article}
\usepackage[dvips]{graphics}
\usepackage{color}
\definecolor{gold}{rgb}{0.85,0.66,0}
\definecolor{dblue}{rgb}{0,0,0.8}
\begin{document}
\onecolumn
\begin{center}
{\bf{\Large {\textcolor{gold}{Quantum transport in honeycomb lattice 
ribbons with zigzag edges: A theoretical study}}}}\\
~\\
{\textcolor{dblue}{Santanu K. Maiti}}$^{1,2}$,\footnote{{\bf
Corresponding Author}: Santanu K. Maiti \\
$~$\hspace {0.45cm} Electronic mail: santanu.maiti@saha.ac.in} \\ 
~\\
{\em $^1$Theoretical Condensed Matter Physics Division,
Saha Institute of Nuclear Physics, \\
1/AF, Bidhannagar, Kolkata-700 064, India \\
$^2$Department of Physics, Narasinha Dutt College,
129 Belilious Road, Howrah-711 101, India} \\
~\\
{\bf Abstract}
\end{center}
We explore electron transport properties in honeycomb lattice ribbons 
with zigzag edges coupled to two semi-infinite one-dimensional metallic 
electrodes. The calculations are based on the tight-binding model and the 
Green's function method, which numerically compute the conductance-energy 
and current-voltage characteristics as functions of the lengths and widths 
of the ribbons. Our numerical results predict that for such a ribbon 
an energy gap always appears in the conductance spectrum across the 
energy $E=0$. With the increase of the size of the ribbon, the gap 
gradually decreases but it never vanishes. This clearly manifests that 
a honeycomb lattice ribbon with zigzag edges always exhibits the 
semiconducting behavior, and it becomes much more clearly visible from 
our presented current-voltage characteristics.

\vskip 1cm
\begin{flushleft}
\end{flushleft}
\vskip 4in

\newpage
\twocolumn

\section{Introduction}

The electronic transport in graphene nanoribbons has opened up new 
challenges in nanoelectronics. A graphene nanoribbon (GNR) is a monolayer 
of carbon atoms arranged in a two-dimensional honeycomb lattice 
structure$^{1-4}$ and can be regarded as the basic
building blocks for graphitic materials. Graphene based materials have
potential applications in several branch of nanoelectronics and due
to their special electronic and physical properties they exhibit several 
novel properties like high carrier mobility,$^3$ unconventional 
quantum Hall effect$^5$ and many others. The high carrier mobility 
in graphene demonstrates the idea for fabrication of high speed switching 
devices those have widespread applications in different fields. Recently
GNRs are also used extensively in designing field-effect transistors and
this idea has been predicted in some recent nice papers.$^{6-8}$ It has 
many potential applications and provides a huge interest in 
the community of nanoelectronics device research. Not only that, GNRs can 
be used to construct MOSFETs which perform much better than conventional 
Si MOSFETs. In other experiment$^9$ it has been proposed that a 
narrow strip of graphene, the so-called a graphene nanoribbon, exhibits 
semiconducting behavior due to its edge effects, unlike carbon nanotubes 
of larger sizes which are mixtures of both metallic and semiconducting 
materials. The reason is that, in a narrow 
graphene sheet a band gap appears across the energy $E=0$, while the gap 
gradually disappears with the increase of the size of the ribbon. It
reveals a transformation from the semiconducting to the metallic material.
This phenomenon has been studied in detail in a very recent theoretical 
work by the same author of this paper.$^{10}$ The situation becomes 
quite different for the case of honeycomb lattice ribbons with zigzag 
edges. Our study predicts that for a ribbon with zigzag edges always 
there exists an energy gap in the conductance spectrum across the energy 
$E=0$. The gap gradually decreases with the increase of the size of the 
ribbon (as expected), but it never vanishes even for much larger ribbons. 
This clearly reveals that a honeycomb lattice ribbon with zigzag edges 
always exhibits the semiconducting behavior, unlike the lattice ribbons 
with armchair edges where both the semiconducting and the metallic phases 
are observed by tuning the size of a ribbon. All these edge effects in 
graphene ribbons provide many key informations in designing 
nanoelectronic devices.

The aim of the present paper is to provide a qualitative study of electron 
transport in honeycomb lattice ribbons with zigzag edges attached to two 
semi-infinite one-dimensional metallic electrodes (see Fig.~\ref{zigzag}). 
The theoretical description of electron transport in a bridge system has 
been developed based on the pioneering work of Aviram and Ratner.$^{11}$ 
Later, many significant experiments$^{12-13}$ have been done in several 
bridge systems to understand the basic mechanisms underlying the electron 
transport. Though in literature many theoretical$^{14-25}$ as well as 
experimental papers$^{12-13}$ on electron transport 
are available, yet lot of controversies are still present between the theory 
and experiment even today. Several controlling factors are there which can 
regulate the electron transport in a conducting bridge significantly, and 
all these factors have to be taken into account properly to understand the 
transport mechanism. For our illustrative purposes, here we mention some 
of these issues. \\
\noindent
(i) The geometry of the conducting material between the two electrodes 
itself is an important issue to control the electron transmission which
has been described quite elaborately by Ernzerhof {\em et al.}$^{26}$ 
through some model calculations. \\
\noindent
(ii) The coupling of the bridging material to the electrodes significantly
controls the current amplitude across any bridge system.$^{27}$ \\
\noindent
(iii) The quantum interference effect$^{27-31}$ of electron waves passing 
through different arms of the bridging material becomes the most 
significant issue. \\
\noindent
(iv) The dynamical fluctuation in 
the small-scale devices is another important factor which plays an active 
role and can be manifested through the measurement of {\em shot 
noise},$^{32-33}$ a direct consequence of the quantization of 
charge. 

In addition to these, several other factors of the Hamiltonian that describe 
a system also provide important effects in the determination of the current 
across a bridge system.

Here we adopt a simple tight-binding model to describe the system and all 
the calculations are performed numerically. We narrate the conductance-energy 
and current-voltage characteristics as functions of the lengths and widths of
ribbons. Our results clearly predicts that a honeycomb lattice ribbon with
zigzag edges always shows the semiconducting nature irrespective of its
length and width.

The paper is arranged in this way. Following the introduction 
(Section $1$), in Section $2$, we present the model and the theoretical 
formulations for our calculations. Section $3$ discusses the significant 
results, and the summary of this work is available in Section $4$.

\section{Model and synopsis of the theoretical background}

\begin{figure}[ht]
{\centering \resizebox*{7.5cm}{4.8cm}{\includegraphics{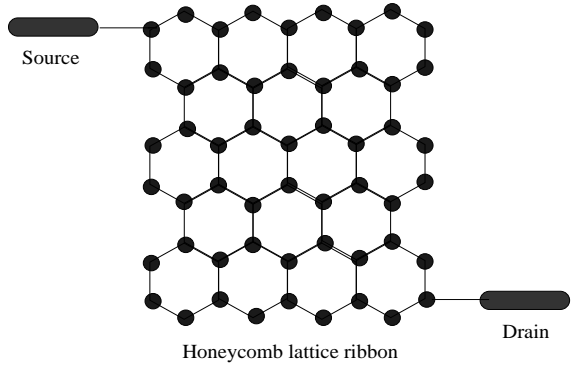}}\par}
\caption{(Color online). Schematic view of a honeycomb lattice ribbon
with zigzag edges attached to two semi-infinite one-dimensional metallic
electrodes, viz, source and drain. Filled circles correspond to the
position of the atomic sites.}
\label{zigzag}
\end{figure}

Let us start with Fig.~\ref{zigzag}, where a honeycomb lattice ribbon with
zigzag edges is attached to two semi-infinite one-dimensional metallic 
electrodes, viz, source and drain. As the electron transport properties 
are significantly
influenced by the quantum interference effects, throughout the study, we 
contact the electrodes at the two extreme ends of nanoribbons (see 
Fig.~\ref{zigzag}) to make these interference effects uniform.
\vskip 0.05in
\noindent
{\em Calculation of the conductance:}
\vskip 0.05in
We use the Landauer conductance formula$^{34-35}$ to calculate
the conductance $g$ of the ribbon. At very low temperature and
bias voltage it ($g$) can be presented in the form,
\begin{equation}
g=\frac{2e^2}{h} T
\label{equ1}
\end{equation}
where $T$ gives the transmission probability of an electron in the ribbon
which can be expressed in terms of the Green's function of the ribbon and 
its coupling to the two electrodes by the relation,$^{34-35}$
\begin{equation}
T={\mbox{Tr}}\left[\Gamma_S G_{rib}^r \Gamma_D G_{rib}^a\right]
\label{equ2}
\end{equation}
where $G_{rib}^r$ and $G_{rib}^a$ are respectively the retarded and advanced
Green's functions of the ribbon including the effects of the electrodes.
The parameters $\Gamma_S$ and $\Gamma_D$ describe the coupling of the
ribbon to the source and drain respectively, and they can be defined in
terms of their self-energies. For the complete system i.e., the ribbon, 
source and drain, the Green's function is defined as,
\begin{equation}
G=\left(\mathcal{E}-H\right)^{-1}
\label{equ3}
\end{equation}
where $\mathcal{E}=E+i\delta$. $E$ is the injecting energy of the source 
electron and $\delta$ gives an infinitesimal imaginary part to $\mathcal{E}$. 
To Evaluate
this Green's function, the inversion of an infinite matrix is needed since
the complete system consists of the finite ribbon and the two semi-infinite 
electrodes. However, the entire system can be partitioned into sub-matrices 
corresponding to the individual sub-systems and the Green's function for 
the ribbon can be effectively written as,
\begin{equation}
G_{rib}=\left(\mathcal{E}-H_{rib}-\Sigma_S-\Sigma_D\right)^{-1}
\label{equ4}
\end{equation}
where $H_{rib}$ is the Hamiltonian of the ribbon which can be written in 
the tight-binding model within the non-interacting picture like,
\begin{eqnarray}
H_{rib}=\sum_i \epsilon_i c_i^{\dagger} c_i + \sum_{<ij>} t 
\left(c_i^{\dagger} c_j + c_j^{\dagger} c_i\right)
\label{equ5}
\end{eqnarray}
In the above Hamiltonian ($H_{rib}$), $\epsilon_i$'s are the site energies, 
$c_i^{\dagger}$ ($c_i$) is the creation (annihilation) operator of an 
electron at the site $i$ and $t$ is the nearest-neighbor hopping integral.
A similar kind of tight-binding Hamiltonian is also used to describe the
two semi-infinite
one-dimensional perfect electrodes where the Hamiltonian is parametrized
by constant on-site potential $\epsilon_0$ and nearest-neighbor hopping
integral $t_0$. In Eq.~\ref{equ4}, $\Sigma_S=h_{S-rib}^{\dagger}g_S 
h_{S-rib}$ and $\Sigma_D=h_{D-rib} g_D h_{D-rib}^{\dagger}$ are the 
self-energy operators due to the two electrodes, where $g_S$ and $g_D$ 
correspond to the Green's functions of the source and drain respectively. 
$h_{S-rib}$ and $h_{D-rib}$ are the coupling matrices and they will be 
non-zero only for the adjacent points of the ribbon, and the electrodes 
respectively. The matrices
$\Gamma_S$ and $\Gamma_D$ can be calculated through the expression,
\begin{equation}
\Gamma_{S(D)}=i\left[\Sigma_{S(D)}^r-\Sigma_{S(D)}^a\right]
\label{equ6}
\end{equation}
where $\Sigma_{S(D)}^r$ and $\Sigma_{S(D)}^a$ are the retarded and advanced
self-energies respectively, and they are conjugate with each other.
These self-energies can be written as,$^{36}$
\begin{equation}
\Sigma_{S(D)}^r=\Lambda_{S(D)}-i \Delta_{S(D)}
\label{equ7}
\end{equation}
where $\Lambda_{S(D)}$ are the real parts of the self-energies which
correspond to the shift of the energy eigenvalues of the ribbon and
the imaginary parts $\Delta_{S(D)}$ of the self-energies represent the
broadening of these energy levels. Since this broadening is much larger 
than the thermal broadening, we restrict our all calculations only
at absolute zero temperature. All the information about the 
ribbon-to-electrode coupling are included into these two self-energies.
\vskip 0.05in
\noindent
{\em Calculation of the current:}
\vskip 0.05in
The current passing across the ribbon can be depicted as a single-electron
scattering process between the two reservoirs of charge carriers. The
current $I$ can be computed as a function of the applied bias voltage $V$
through the relation,$^{34}$
\begin{equation}
I(V)=\frac{2e}{h} \int \limits_{E_F-eV/2}^{E_F+eV/2} T(E) ~dE
\label{equ8}
\end{equation}
where $E_F$ is the equilibrium Fermi energy. Here we make a realistic
assumption that the entire voltage is dropped across the ribbon-electrode
interfaces, and it is examined that under such an assumption the $I$-$V$
characteristics do not change their qualitative features. This assumption
is based on the fact that, the electric field inside the ribbon especially 
for narrow ribbons seems to have a minimal effect on the conductance-voltage 
characteristics. On the other hand, for quite larger ribbons and high bias 
voltages the electric field inside the ribbon may play a more significant 
role depending on the internal structure and size of the ribbon,$^{36}$ 
but the effect becomes too small.

\section{Numerical results and discussion}

To have a better insight in the present problem i.e., the dependence of 
the electron transport on the lengths and widths of ribbons, we focus our
attention only on the perfect systems rather than any disordered one. To
get these perfect systems we fix the on-site energies $\epsilon_i=0$ for 
all the sites $i$ of the honeycomb lattice ribbons. The values of the 
other parameters are taken as follow. The nearest-neighbor 
hopping integral $t$ in the ribbon is set to $1$, the on-site energy 
$\epsilon_0$ and the hopping integral $t_0$ for the two electrodes are 
fixed to $0$ and $1$ respectively. The parameters $\tau_S$ and $\tau_D$ 
are set as $0.75$, where they correspond to the hopping strengths of the 
ribbon to the source and drain respectively. In addition to these, to 
describe the size of a ribbon we introduce two other parameters $N$ and 
$M$ where they correspond to the width and length of the ribbon respectively. 
Thus, for example, a nanoribbon with $N=1$ and $M=3$ represents a linear 
chain of three hexagons. Hence the parameter $M$ determines the total number 
of hexagons in a single chain. Following this rule, a nanoribbon with $N=3$ 
\begin{figure}[ht]
{\centering \resizebox*{7.75cm}{10cm}{\includegraphics{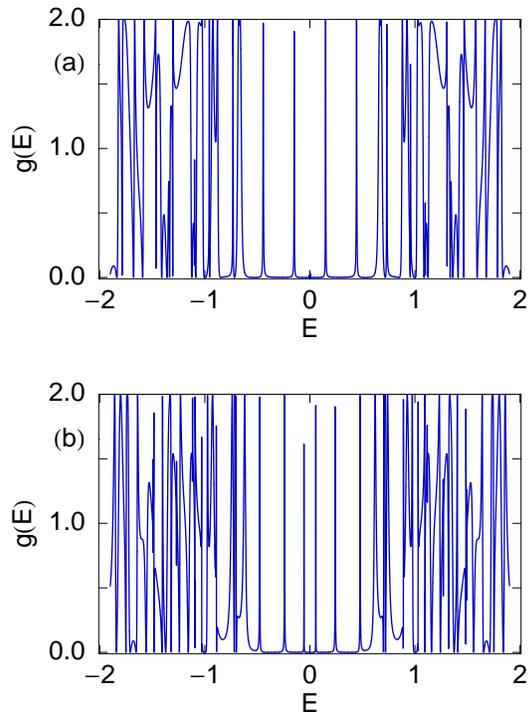}}\par}
\caption{(Color online). Conductance $g$ as a function of the energy $E$
for some lattice ribbons with fixed width $N=3$ and varying lengths where
(a) $M=6$ and (b) $M=8$.}
\label{cond1}
\end{figure}
and $M=4$ corresponds to three linear chains attached side by side (see 
Fig.~\ref{zigzag}) where each chain contains four hexagons. For simplicity, 
throughout our study we set the Fermi energy $E_F=0$ and choose the units 
where $c=e=h=1$.

Let us begin our discussion with the variation of the conductance $g$ as
a function of the energy $E$. In Fig.~\ref{cond1} we plot the 
conductance-energy characteristics for some typical lattice ribbons
with fixed width $N=3$ and varying lengths where (a) and (b) correspond
to the lengths $M=6$ and $8$ respectively. The sharp resonant peaks
in the conductance spectra are observed almost for all energies, while
for some other energies either the conductance $g$ gets much small value 
or drops to zero. At the resonances the conductance gets the value $2$,
and therefore, the transmission probability $T$ becomes unity since the 
relation $g=2T$ follows from the Landauer conductance formula, 
Eq.~\ref{equ1}, with $e=h=1$. Now the reduction of the transmission 
probability ($T<1$) for some particular energies can be explained by
considering the quantum interference effects of the electronic waves 
passing through the different arms of the ribbon. The interpretation
is as follow.
\begin{figure}[ht]
{\centering \resizebox*{7.75cm}{10cm}{\includegraphics{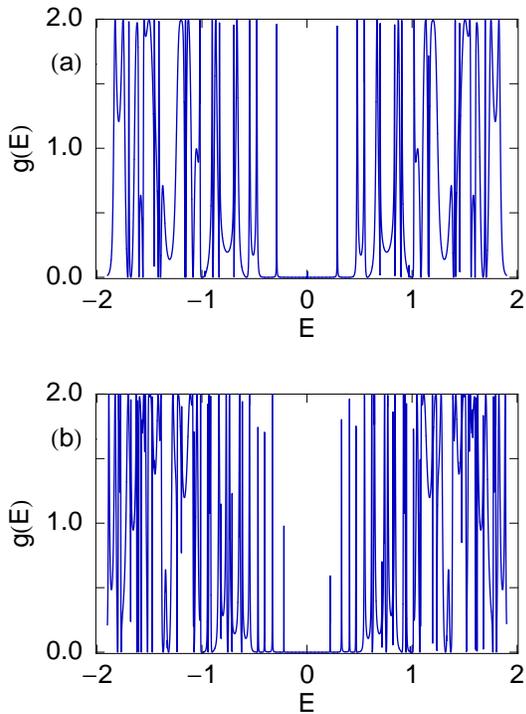}}\par}
\caption{(Color online). Conductance $g$ as a function of the energy $E$
for some lattice ribbons with fixed length $M=4$ and varying widths where
(a) $N=5$ and (b) $N=10$.}
\label{cond2}
\end{figure}
During the motion of the electrons from the source to drain through the 
lattice ribbon, the electron waves propagating along the different possible 
pathways can get a phase shift among themselves, according to the result 
of quantum interference. Therefore, the probability amplitude of getting 
an electron across the ribbon either becomes strengthened or weakened. 
This causes the transmittance cancellations and provides anti-resonances 
in the conductance spectrum. Thus it can be emphasized that the electron
transmission is strongly affected by the quantum interference effects,
and hence the ribbon to electrode interface structure. Now all these
resonant peaks are associated with the energy eigenvalues of the ribbon, 
and accordingly, we can say that the conductance spectrum manifests itself 
the electronic structure of the ribbon. Due to the large number of energy 
levels, associated with the size of the ribbons, the resonant peaks almost 
overlap with each other and form quasi-continuous spectra in the $g$-$E$ 
characteristics. The most significant feature observed
\begin{figure}[ht]
{\centering \resizebox*{7.75cm}{5cm}{\includegraphics{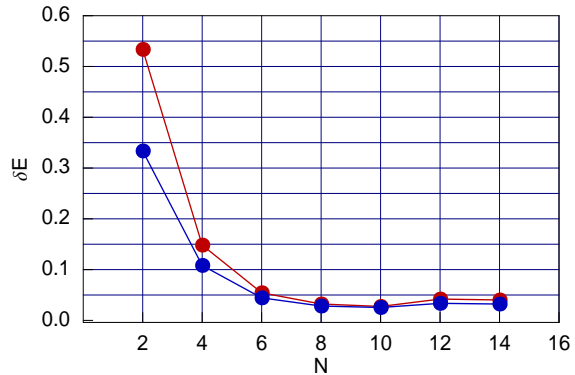}}\par}
\caption{(Color online). Variation of the central energy gap $\delta E$ as
a function of the width $N$ for some lattice ribbons with fixed lengths $M$.
The red and blue curves correspond to $M=2$ and $4$ respectively.}
\label{gap}
\end{figure}
from the spectra, given in Fig.~\ref{cond1}, is that a central energy gap 
($\delta E$) appears across the energy $E=0$. With the increase of the 
length of the ribbon some more resonant peaks appear around $E=0$, and 
accordingly, the gap decreases which is clearly observed from 
Figs.~\ref{cond1}(a) and (b). Thus for a fixed width, the central gap 
can be controlled by tuning the length of the ribbon.

In the same fashion, to characterize the dependence of the electron transport
on the widths of the ribbons for a fixed length, in Fig.~\ref{cond2} we
plot the results for some typical lattice ribbons considering the length
$M=4$. The spectra shown in Figs.~\ref{cond2}(a) and (b) correspond to the
ribbons with widths $N=5$ and $10$ respectively. Quite similar to the 
above case, here also a gap appears across $E=0$ and it decreases 
with the increase of the width of the ribbon though the reduction is much 
small. Thus from the results described in Figs.~\ref{cond1} and \ref{cond2}
we can emphasize that the width of the central energy gap always decreases 
with the increase of the size (length and width) of the ribbon.

To illustrate the dependence of the gap $\delta E$ with other system
sizes, in Fig.~\ref{gap}, we display the variation of $\delta E$
as a function of the width $N$ for some typical lattice ribbons with fixed
lengths. The red and blue curves correspond to the lengths $M=2$ and $4$
respectively. Quite interestingly we see that, the gap $\delta E$ gradually 
decreases with the increase of the width $N$, and beyond a certain value 
of $N$, the rate of decrease of this gap becomes much small and eventually
it ($\delta E$) becomes almost a constant. Quite similar feature is also 
observed if we plot the variation of the energy gap as a function of the 
length $M$ keeping the width $N$ as a constant, and due to the obvious 
reason we do not plot these results further in the present description.
\begin{figure}[ht]
{\centering \resizebox*{7.75cm}{10cm}{\includegraphics{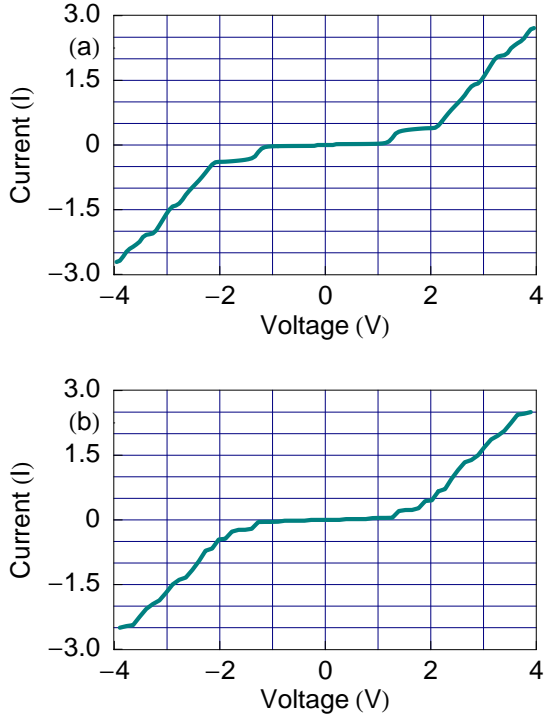}}\par}
\caption{(Color online). Current $I$ as a function of the bias voltage $V$
for some lattice ribbons with fixed width $N=3$ and varying lengths where
(a) $M=3$ and (b) $M=6$.}
\label{current1}
\end{figure}
These results provide us an important signature which concern with the
variation of the energy gap by tuning the size of the ribbon, and we
can emphasize that a honeycomb lattice ribbon with zigzag edges always
exhibits the semiconducting (finite energy gap) behavior.

\begin{figure}[ht]
{\centering \resizebox*{7.75cm}{10cm}{\includegraphics{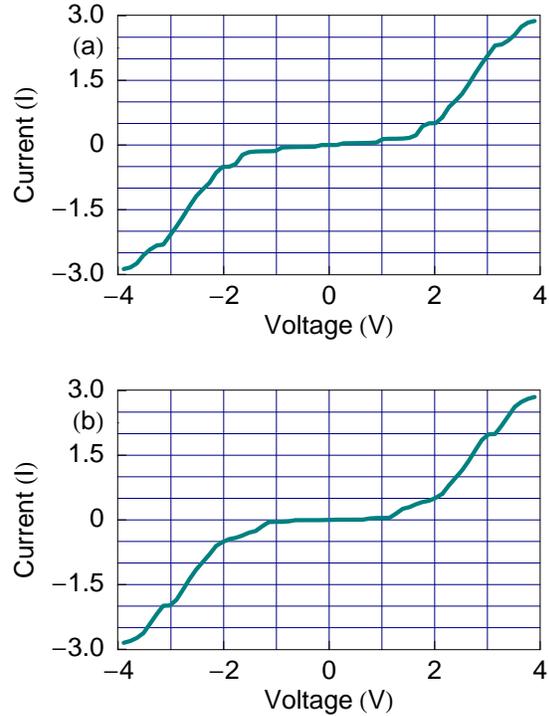}}\par}
\caption{(Color online). Current $I$ as a function of the bias voltage $V$
for some lattice ribbons with fixed length $M=4$ and varying widths where
(a) $N=2$ and (b) $N=3$.}
\label{current2}
\end{figure}

All these basic features of electron transfer can be much more clearly 
explained from our investigation of the current-voltage ($I$-$V$) 
characteristics rather than the conductance-energy spectra. The current 
$I$ is determined from the integration procedure of the transmission 
function ($T$) (see Eq.~\ref{equ8}), where the function $T$ varies 
exactly similar to the conductance spectra, differ only in magnitude 
by a factor $2$, since the relation $g=2T$ holds from the Landauer 
conductance formula (Eq.~\ref{equ1}). As an illustration, in 
Fig.~\ref{current1}, we present the current-voltage ($I$-$V$) 
characteristics for some lattice ribbons with fixed width $N=3$ and 
varying lengths where (a) and (b) correspond to the lengths $M=3$ and 
$6$ respectively. In the same footing, in Fig.~\ref{current2}, we plot 
the variation of the current $I$ as a function of the bias voltage 
$V$ for some typical lattice
ribbons keeping the length as fixed ($M=4$) and vary the widths, where (a)
and (b) represent the ribbons with widths $N=2$ and $3$ respectively.
The sharpness in the $I$-$V$ characteristics and the current amplitude 
solely depend on the coupling strengths of the ribbon to the side attached
electrodes, viz, source and drain. It is observed that, in the limit of 
weak coupling, defined by the condition $\tau_{S(D)} << t$, current shows 
staircase like structure with sharp steps. While, in the strong coupling 
limit, described by the condition $\tau_{S(D)} \sim t$, current varies quite
continuously with the bias voltage $V$ and achieves large current amplitude
compared to the weak-coupling limit. All these coupling effects have
already been explained in many theoretical as well as experimental papers 
in the literature. The key feature observed from these $I$-$V$ characteristics
is that for all such ribbons the electron starts to conduct beyond some 
finite bias voltage, the so-called threshold bias voltage $V_{th}$, and it 
decreases very slowly with the change of the size of the ribbon. Our
study reveals that the threshold bias voltage never drops to zero, even
for much larger systems, and accordingly, we can predict that a honeycomb 
lattice ribbon with zigzag edges exhibits only the semiconducting behavior.

\section{Concluding remarks}

To summarize, we have addressed electron transport properties in honeycomb 
lattice ribbons with zigzag edges attached to two semi-infinite 
one-dimensional metallic electrodes within the tight-binding framework. 
We have numerically computed the conductance-energy and current-voltage
characteristics concerning the dependence on the lengths and widths of
the ribbons. The results have predicted that for such ribbons i.e., 
ribbons with zigzag edges a central energy gap always exists across the 
energy $E=0$ in the conductance spectrum. The gap decreases gradually 
with the increase of the size of the ribbon but it never vanishes. This 
phenomenon clearly manifests that a honeycomb lattice ribbon with zigzag 
edges always shows semiconducting nature, unlike the lattice ribbons 
with armchair edges where both the semiconducting and the metallic 
phases are observed by controlling the size of the ribbon. This 
semiconducting behavior of the lattice ribbons with zigzag edges has
been much more clearly addressed from our presented current-voltage
characteristics. It has been observed that the current starts to appear
beyond some finite bias voltage i.e., the threshold bias voltage $V_{th}$
has a non-zero value. This $V_{th}$ doesn't change appreciably with 
the increase of the size of the ribbon and we have also examined that 
the threshold bias voltage never reduces to zero even for much larger 
systems, which predicts the semiconducting nature only. 

This is our first step to describe how the electron transport properties
in honeycomb lattice ribbons with zigzag edges depends on the size of the 
ribbons. We have made several realistic assumptions by ignoring the
effects of the electron-electron correlation, disorder, interaction with
a substrate, temperature, finite width of the electrodes, boundary of the
ribbons, etc. Here we discuss very briefly about these approximations.
The inclusion of the electron-electron correlation in the present model
is a major challenge to us, since over the last few years people have 
studied a lot to incorporate this effect, but no such proper theory has
yet been developed. In this work, we have presented all the results only
for the ordered systems. But in real samples, the presence of impurities
will affect the electronic structure and hence the transport properties.
Beside these, in experiments, the graphene nanoribbon is deposited on an
insulating substrate which has also not been included in our present study,
and, it has been observed from first-principles calculations that the
effect of the substrate is too week.$^{37}$ The effect of the
temperature has already been pointed out earlier, and, it has been
examined that the presented results will not change significantly even
at finite temperature, since the broadening of the energy levels of the
ribbon due to its coupling with the electrodes will be much larger than
that of the thermal broadening.$^{34}$ The other important assumption
is that here we have chosen the linear chains instead of wider leads,
since we are mainly interested about the basic physics of the ribbon.
Though the results presented here change with the increase of the
thickness of the leads, but all the basic features remain quite invariant.
The effect of the boundary is also an important issue in this context.
Here we have considered only the perfect geometry of the nannoribbons.
Several interesting features will be observed for the nanoconstrictions
with different shapes$^{38}$ like, square-shaped, wedge-shaped
nanoconstrictions, etc. Finally, we would like to say that we need
further study in such systems by incorporating all these effects.

\end{document}